# AI-MACHINE LEARNING-ENABLED TOKAMAK DIGITAL TWIN

William Tang, Eliot Feibush, Ge Dong, Noah Borthwick, Apollo Lee, Juan-Felipe Gomez
Princeton University/Princeton Plasma Physics Laboratory, Princeton, NJ, USA
Email: wtang@princeton.edu

Tom Gibbs, John Stone, Peter Messmer, Jack Wells
NVIDIA Corporation, Santa Clara, CA, USA

Xishuo Wei, Zhihong Lin
University of California Irvine, Irvine, California USA

**Abstract**

In addressing DOE's April, 2022 announcement of a Bold Decadal Vision for delivering a Fusion Pilot Plant by 2035 [1], associated software tools need to be developed for the integration of "real world engineering and supply chain data" with advanced science models that are accelerated with Machine Learning [2]. An associated R&D effort has been introduced here with promising early progress on the delivery of a realistic "Digital Twin Tokamak" that has benefited from accelerated advances by Princeton University's AI/Deep Learning innovative near-real-time simulators accompanied by technological capabilities from the NVIDIA Omniverse™ — an open computing platform for building and operating applications that connect with leading scientific computing visualization software. Working with the CAD files for the GA/DIII-D tokamak including equilibrium evolution as an exemplar tokamak application using Omniverse, the Princeton-NVIDIA collaboration has integrated modern AI/HPC-enabled near-real-time kinetic dynamics to connect and accelerate state-of-the-art, synthetic, HPC simulators to model fusion devices and control systems. The overarching goal is to deliver an interactive scientific digital twin of an advanced MFE tokamak that enables near-real-time simulation workflows built with Omniverse to eventually help open doors to new capabilities for generating clean power for a better future.

1.  INTRODUCTION

To help realize the US DOE Bold Decadal Vision [1] modern tools which integrate engineering and supply chain data with advanced science models that are accelerated with Machine Learning (ML) will be required [2]. Accordingly, we have initiated and obtained encouraging results from a research & development (R&D) effort with progress toward the delivery of a realistic "Digital Twin Tokamak" that has been accelerated by advances in artificial intelligence and deep learning (AI/Deep Learning) for providing innovative real-time simulators accompanied by technological capabilities from NVIDIA Omniverse™ — an open computing platform for building and operating applications capable of connecting to leading scientific computing visualization software. Working with the CAD files for the GA/DIII-D tokamak [3] as an exemplar tokomak application using Omnivere, we have incorporated recent near-real-time simulators enabled by AI/HPC (high-performance computing) to provide a viable approach to representing near-real-time kinetic plasma dynamics.

The Princeton-NVIDIA collaboration has connected and accelerated state-of-the-art, synthetic, near-real-time HPC simulators to model fusion devices and control systems targeting the goal of improving the operation of tokamak experiments leading to a new commercially viable clean-energy source [1]. This is crucially important for research scientists and engineers — as well as being a major current target for government organizations. The ultimate outcome is to deliver an interactive scientific digital twin of a fusion device that enables real-time simulation workflows built to open doors to new capabilities for generating clean power for a better future.

2. BACKGROUND

Fusion energy science simulations are generating increasingly large datasets, and modern visualization methods have been key for exploring, verifying, and communicating associated results. As both experimentally-measured and simulated data size continues to rapidly increase, computational scientists and data analysis software engineers need to identify and deploy the best data selection methods. In the plasma physics/fusion energy science area it is important to collect and curate data that include locations of specific magnetic value, such as flux tubes, and working with challenging 3-D datasets acquired from experiments with a 3-D compute grid. As a complex visualization capability, Omniverse can deliver real-time videos capable of greatly contributing to improved visual communication and higher audience engagement. An important component of creating an animated video is planning the path the camera takes as it tours the realistic model. While existing approaches rely on key-frame views, Omniverse-like tools would be required for tasks such as following 3-D curves integrated with the data selection. From the Fusion mission perspective, a near real-time physics-based simulator is clearly needed for predictive guidance since (i) observational data is unavailable for designed devices not yet operational with ITER being the prime example; and (ii) experimentally-validated 1st principles HPC codes (such as the global electromagnetic gyrokinetic toroidal code (GTC) [4] are all far too slow to be useful in near-real time applications. The recent arrival of the synthetic GTC workflow [SGTC] with AI neural network based *surrogates* that have been experimentally validated against the 1st-principles electromagnetic global PIC (particle-in-cell) code GTC have made a significant impact. SGTC has been trained with extensive simulated data from GTC to enable inference prediction of dynamical plasma instability properties on millisecond time scales compatible with actuator operations in a PCS (plasma control system). As reported in the peer-reviewed journal Nuclear Fusion [5], SGTC can *deliver results 5 orders of magnitude faster than GTC runs on HPC Leadership Class Facilities*. It should be noted that as a reduced model surrogate derived from the traditional GTC model, SGTC can examine validation/accuracy considerations by exploring the wide middle ground of speed/accuracy tradeoff with validation tests carried out on the DIII-D tokamak.

3. PRIOR WORK ON DIGITAL TWINS

The concept of a Digital Twin was first proposed as a "Mirror World" by Gelernter [6]. Based on digital models of existing or proposed buildings, users would be able to explore functions and operations within a building through interactive programs. More specific descriptions of Digital Twins in manufacturing were provided by Grieves [7], and a recent comprehensive treatment of the history, definition, and also misconceptions can be found in the publication by Fuller [8]. Motivated by Fusion engineering and design considerations for the KSTAR tokamak, a virtual tokamak platform was introduced by Kwon [9] – a system based on the KSTAR tokamak that includes a CAD model, plasma simulation, and visualization using the Unity game engine.

4. REPRESENTATIVE TOKAMAK DIGITAL TWIN WORKFLOW

The workflow for the Fusion Digital Twin in our paper encompasses -- as illustrated in Fig. 1 -- engineering, experimentation, theory, computation, and visualization. This comprehensive approach begins with data acquired from experimental shots on an operating tokamak (DIII-D) *but can be readily extended to other tokamaks*. The data is analyzed and used as input to a large-scale gyrokinetic simulation carried out by the electromagnetic PIC GTC code. An extensive number of simulation runs carried out on HPC leadership-class supercomputers then provide training data to the machine learning based surrogate (SGTC) simulation. The combined data from the experiment and the simulations is displayed in the context of a detailed CAD model.

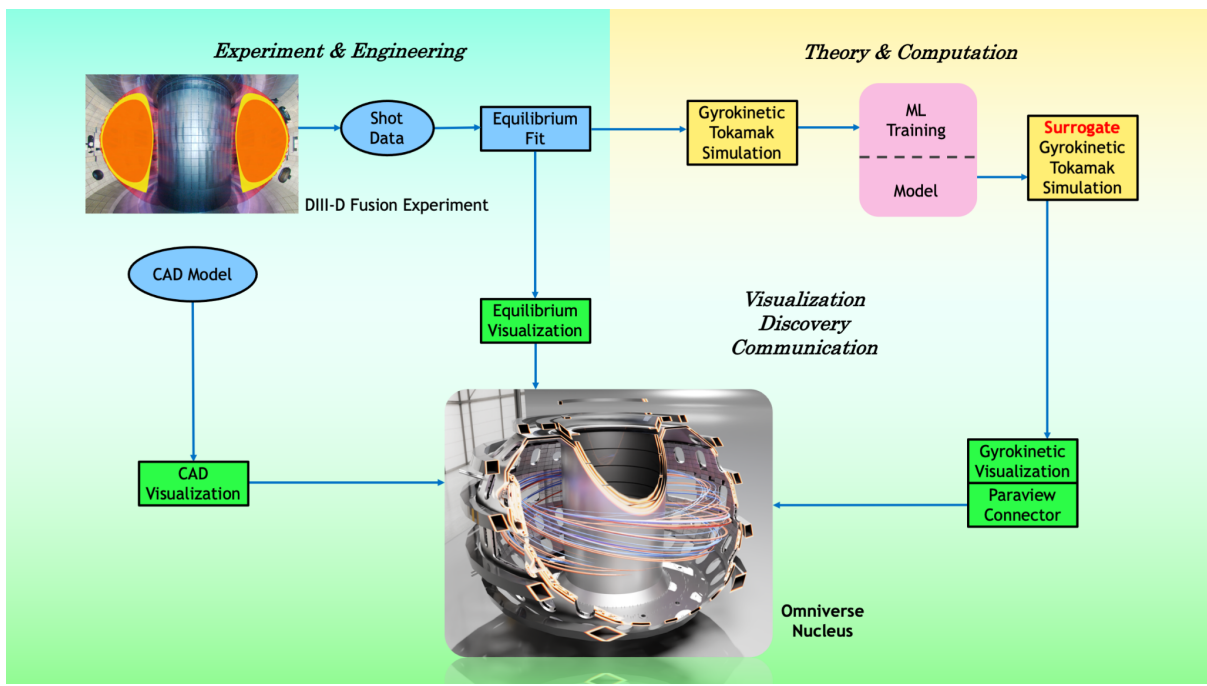

*Fig. 1 The illustrated workflow for the Fusion Digital Twin is an example based on the DIII-D tokamak but can be adapted for other MFE devices of interest.*

The gyrokinetic simulation runs were carried out on a large number of cores of supercomputers (e.g., SUMMIT at ORNL [10]) for many hours provided by DOE INCITE allocations [11]. The output from these runs provide the essential training data required for the machine learning model. It is important to note that this training process runs in advance of pairing the Digital Twin with the actual experiment. The CAD model is static and converted to visualization format as a pre-process. Other modules run fast enough to be paired with a running experiment. The current prototype implementation stores data in files to communicate between modules. Inter-process communication and shared memory will be deployed in future versions to optimize the data flow performance.

**4.1 Experiment data input into Digital Twins**

The experimental data (here from DIII-D) provide the inputs for the Digital Twin inputs that are used in the equilibrium fitting code, EFIT originally described by Lao [12]. Based on magnetic probes, poloidal flux loops, and the Motional Stark Effect (to measure magnetic field lines), EFIT constructs the current profile and plasma shape as part of a plasma equilibrium that evolves during a shot. The profiles and magnetic fields then serve as inputs to the gyrokinetic tokamak code, GTC. Other tokamaks and MFE devices of interest can similarly be studied.

Porting and integrating all the components of the Digital Twin workflow is a software challenge. Some of the code for data preparation is very specialized plasma physics software with many dependent software modules. Porting all the code to a new platform is complicated due to compiler and library dependencies. We are looking to overcome this problem by putting the specialized executable programs into software containers [16, 17] and moving only the containers to the Digital Twin target platform.

**4.2 Gyrokinetic simulation and surrogate development of Digital Twins**

Innovation in the development of improved data-driven and experimentally validated model-based approaches are needed to maximize plasma performance in existing experiments with impact on optimizing operational scenarios. In the present studies it also serves to accelerate progress on Digital Twin development

The deployment of recurrent and convolutional neural networks in Princeton's Deep Learning Code "FRNN" enabled the first adaptable predictive deep learning (DL) model for carrying out efficient "transfer learning" with accurate predictions of disruptive events across different tokamak devices [13]. The demonstrated successful validation of FRNN software on a huge observational FES database provided strong evidence that DL approaches using large scale supercomputers could predict disruptions with compelling accuracy. More recent publications have further shown that this AI/DL capability can provide not only the "disruption score," as an indicator of the probability of an imminent disruption but also a "sensitivity score" in real-time to indicate the underlying reasons for the predicted disruption – i.e., "explainable AI." [14].

Ongoing efforts address efficient integration of the evolving improved version of the AI/Deep Learning FRNN prediction and control software into the real-time DIII-D PCS. This kind of systematic statistical analysis has continued to exhibit compelling predictive capability using archived data, as described in the NATURE paper [13]. The newest version of the FRNN inference engine has demonstrated "real-time" functionality in the DIII-D PCS with characteristic execution times of less than 1.7 msec. – and thereby engage in systematic testing of an array of actuators. In future investigations, this will be guided by a proper implementation of FRNN within a "proximity control architecture" designed to identify and predict the major causes of disruptions [15].

The overarching aim here is to demonstrate that real-time modification of the plasma state can lead to a more favorable thermodynamic one to avoid or delay the onset of disruptions. In order to systematically progress toward our stated goal, a key AI/DL (artificial intelligence/deep learning) software challenge is to build a modern high-performance computing enabled "synthetic plasma simulator" capable of carrying out HPC-driven real-time plasma control applications. This involves development of a deep learning framework to train the surrogate model for a first principles-based instability analysis simulator ("SGTC") derived from the global gyrokinetic code GTC [5]. The role of SGTC is to provide accurate and detailed plasma instability information from a real-time AI-based simulator capability to complement the deep learning prediction and control from experimentally measured signals. A major data management challenge will involve the collection and curation of the expected huge volume of measured and simulated data to enable scientific discoveries.

Surrogates are key for providing detailed information to Digital Twins to realistically enable a plasma control system to help improve disruption avoidance in near-real-time to optimize plasma performance. In particular, the application of AI/DL methods for real-time prediction and control has recently been demonstrably advanced with the introduction of a surrogate model/HPC simulator ("SGTC") [5]. SGTC models satisfy PCS compatibility requirements and deliver inference times on order of milliseconds (ms) and – most impressively -- can demonstrably deliver results *5 orders of magnitude faster than the validated first-principles-based global particle-in-cell GTC code runs on advanced leadership computing HPC systems.* The gyrokinetic simulation code GTC and its machine learning based surrogate SGTC provide the physical basis for the Digital Twin. Using the large database generated from GTC simulations, SGTC can give accurate stability information with about 1ms inference time

These capabilities are now leading to exciting avenues for moving from passive prediction to active control as part of the Tokamak Digital Twin campaign to help optimize the design for a first-of-a-kind fusion pilot plant (FPP).

SGTC software has been currently implemented to predict the internal kink mode instability and introduced a novel and realistic AI-enabled model for the q-profile reconstruction. The n=1 mode (including the kink mode and the neoclassical tearing mode (NTM) are considered to have close relation to fast disruptions in tokamaks. Based on the GTC simulation data of about 5,000 DIII-D shots with strong n=1 signals, the SGTC surrogate model has been trained to predict the linear stability and 2D mode structure of the kink mode [5]. The AUC value of the stability prediction ROC curve is 0.945, indicating the mode structure can be successfully predicted as illustrated in Fig. 2 for 85% of the test dataset. The fast inference speed allows us to gain the information of the kink mode for a long time period of one shot (as seen in Fig. 3), which is clearly an unrealistic task for a traditional first-principles-based code. In addition, the SGTC surrogate can be integrated with the FRNN framework to improve the capability of disruption prediction by providing more accurate kink mode information. GTC/SGTC also provides the grid mapping from the 2D mode structure to the 3D space for later visualization.

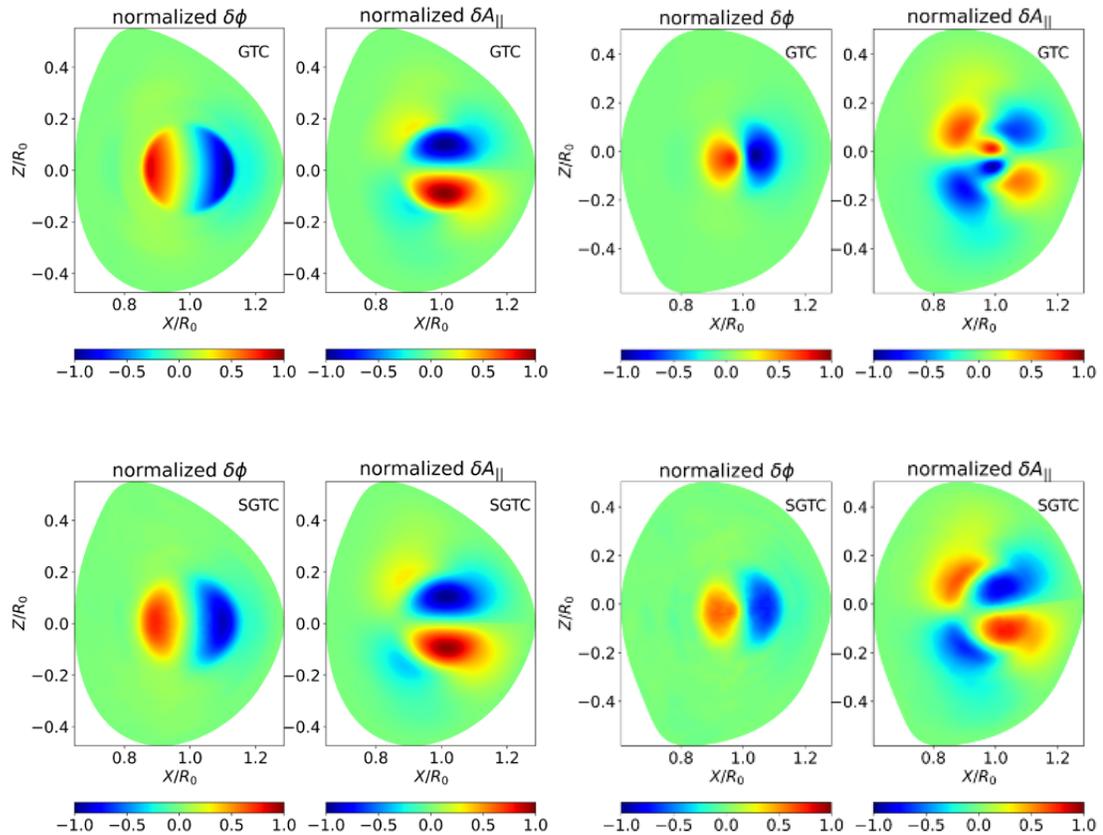

*Fig. 2 The kink mode structure from GTC simulation (upper panel) and the surrogate SGTC prediction (lower panel) for two different shots.*

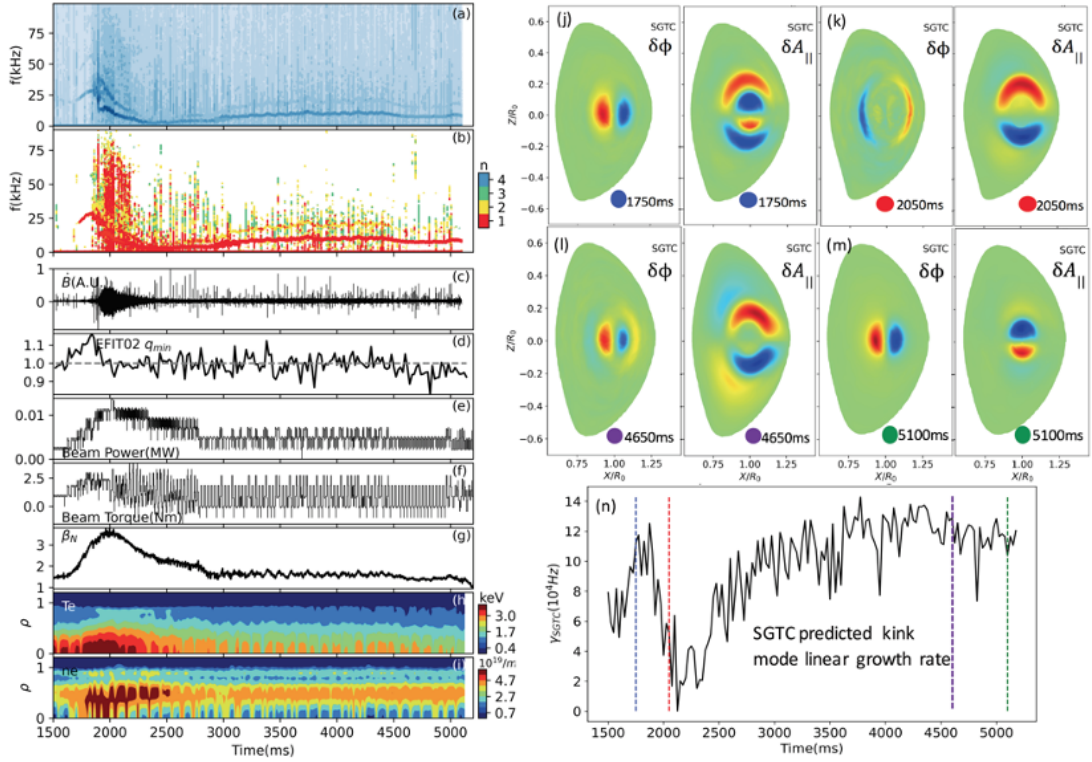

*Fig. 3 The surrogate SGTC kink mode prediction for DIII-D shot #141216 from 1500 ms to 5000 ms.*

As we carried out a large number of kink mode simulations, we found that the quality of q-profile reconstruction is critical for accurate GTC simulations and associated SGTC predictions. Our equilibrium data including q-profile comes from the widely used reconstruction code EFIT. The Motional Stark Effect (MSE), which reflects the pitch angle of magnetic field lines in the tokamak, is an important constraint for accurate q-profile reconstruction. One of the most important upgrades from the so-called "EFIT01" to the "EFIT02" is the inclusion of this MSE constraint. However, MSE can be unavailable for experiments due to engineering or economical constraints. Therefore, we have developed the SGTC q-profile reconstruction model (SGTC-QR) to recover the EFIT02 result without directly using the MSE data. The equilibria and fluctuation signals of 12,000 DIII-D shots and the q-profile EFIT02 reconstruction were used as the target prediction for the machine learning algorithm. Our results show that using the experimental signals other than MSE data, SGTC can achieve a similar q-profile result as EFIT02, as shown in Fig. 3. This study shows that the MSE measurement can be correlated inherently with other measurements. The methodology introduced here is significant because replacing the expensive physical diagnostics with machine learning algorithms in the planning and design of fusion pilot plants can be quite important. In future studies, the recovery of other measurements will be studied, and a complete equilibrium reconstruction, including the flux surface shape, will also be investigated.

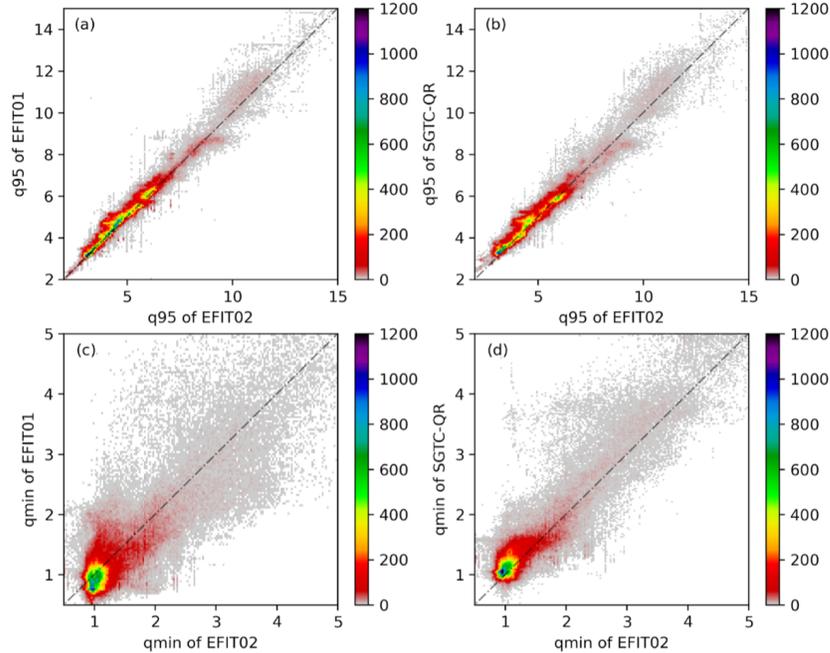

*Fig. 4. Illustration of the SGTC q-profile reconstruction model (SGTC-QR) achievement of recovering the EFIT02 result without directly using the MSE data.*

**4.3 Visualization for Digital Twins**

Omniverse is a platform for interactive display and interaction with large 3-D models, live simulations, and is aptly suited as the visualization component in Digital Twins. It achieves high performance through a combination of on-the-fly optimization of geometry and level of detail, state-of-the-art rendering and AI upscaling and denoising algorithms, and the use of purpose-built ray tracing acceleration features in NVIDIA RTX GPUs. This enables the display of highly detailed CAD models along with 3D representations of the magnetic field and plasma features. Large data geometric meshes can be processed directly by Omniverse and do not have to first be reduced or converted to textures to accommodate the limits of rendering algorithms commonly employed by game engines or similar tools. Recent RTX GPUs provide 24 to 48 GB of memory for storing graphics data. Omniverse can bring in data from multiple sources simultaneously, such as the CAD model, the magnetic field equilibrium, and electrostatic potential from gyrokinetic simulations. Predictions of the kink mode can also be added to the visualization.

The Omniverse USD Composer Section Tool allows users to create cut-away views of the interior. The selection can be interactively positioned by the user or programmatically set by a script for repetitive views. For the DIII-D exemplar study here, the standard equilibrium fitting code EFIT [12] produces an $f(x,y)$ poloidal cross section of the magnetic field. This 2D slice is rotated around the vertical axis of the tokamak to create an axisymmetric reconstruction of the full 3D torus. In the file-based prototype of the Digital Twin the EFIT field is written as a sparse volume in VDB format, and is displayed with the built-in volume rendering capabilities in Omniverse.

The compute grid can be aligned to the xyz Cartesian coordinate system or to spherical coordinates depending on the application. Typically, values are computed at each grid point or each cell with scalar, vector, or higher order values being calculated. Temporal selection of the geometry of an "isovolume" and using it for all time steps would be an effective new approach for exploring an active region. Selected flux tubes of electrostatic potential are shown in a cut-away view of the vacuum vessel in Fig. 1.

The preliminary approach is to model the visualization grid in Visualization Toolkit format, display in Paraview, and subsequently use the Paraview Connector to synchronously stream the data into Omniverse. The filters in Paraview can be utilized to calculate isosurfaces, isovolumes, and selection thresholds. Adapting the selection over time or to analytical conditions can enable advanced exploration methods with associated models benefiting from effective viewing and exploration. Performance optimization can be achieved by initially establishing the geometric grid of the visualization data. For a variable at each time step, e.g. electron density, only the new scalar values need to be computed and sent to the display structure.

Our digital twin studies will help create tools for interactively drawing and previewing camera paths through the environment, including automating the path-plans based on specific data selection techniques. This involves integrating scripting languages to automate assembling movies composed of animated sequences, still images, text, and audio to raise the quality of visual communications.

5. CONCLUSIONS

In contrast with our workflow for the Fusion Digital Twin shown on Fig.1, theory and simulation have often been viewed as an "island" separate from the operation of the experiment and engineering. ML now allows these functions to be bridged – with the Digital Twin using ML to enable the ability to interact with the model as if it were the actual physical device. In future implementation studies of Tokamak Digital Twins targeted plans include the following topics.

For specific physics applications, we plan to develop and run with our UC-Davis collaborators a synthetic ECEI capability with the SGTC output database of gyrokinetic results. This will enable specific MHD "kink and fishbone" instability investigations. The synthetic SGTC-ECEI output (developed and led by our UC-Davis collaborators can be upgraded to be used to provide numerical temperature fluctuation contours with SGTC. In future investigations, the experimental observation results can be directly imported to FRNN for real-time studies of key instability characteristics that can include radial/poloidal envelope information of temperature fluctuations, wavenumbers, and growth rates. These features can be both convenient and highly-relevant for instability prediction and avoidance of dangerous events.

Other stimulating targeted Digital Twin studies for the near future will involve:
- Energetic-Particle/Burning Plasma Dynamics relevant for FPP and ITER with extensive NTM studies;
- Demonstration of Digital Twin adaptability to scenarios including long-pulse KSTAR (Korea); restored TFTR data including reconstructed CAD-file information. Key restored data from TFTR include signals analogous to the AI/Deep Learning disruption studies in our 2019 NATURE paper using the large datasets from the EUROfusion/JET and DIII-D data bases.

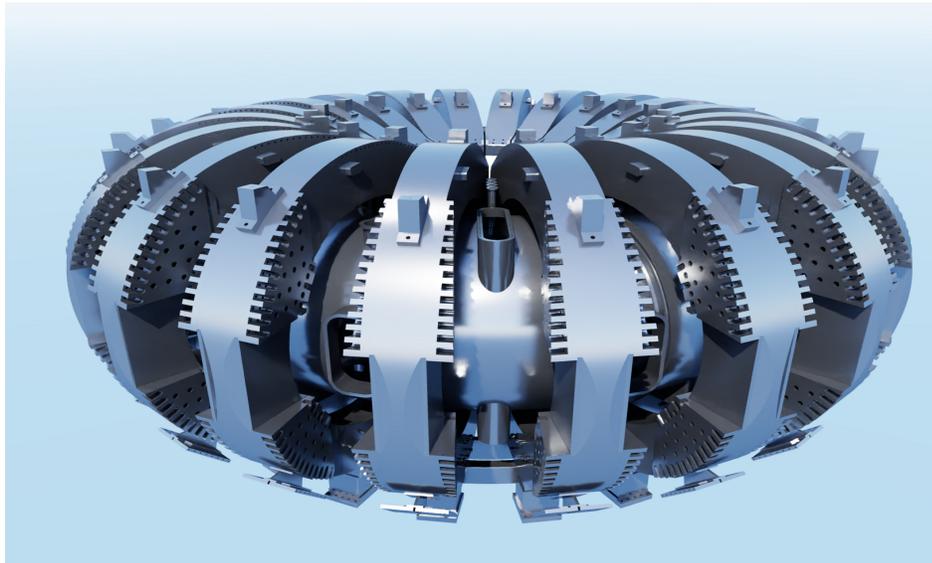

*Fig. 5  Princeton's Tokamak Fusion Test Reactor (TFTR) has been decommissioned but a high-fidelity digital twin would provide an effective testbed for virtual experiments. There would be considerable educational opportunities for scientists and engineers to work with a historical tokamak that is no longer available. A highly realistic rendering of the TFTR vacuum vessel and poloidal field coils has been produced by Omniverse from the CAD model.*

Finally, it is of interest to point out in the context of innovative future models and algorithms enabled by AI with relevance to digital twins, the topic of "generative AI" holds exciting promise for creating new output (text, photos, videos, code, data, 3D renderings, etc.) from the vast amount of data on which they can be trained. Such models can "generate" associated novel content by connecting back to the original training data to make "new predictions."

## REFERENCES

[1] https://www.whitehouse.gov/ostp/news-updates/2022/04/19/readout-of-the-white-house-summit-on-developing-a- bold-decadal-vision-for-commercial-fusion-energy/

[2] Karen Wilcox, Aerospace Engineering, UT Austin: https://www.ted.com/talks/karen_willcox_how_digital_twins_could_help_us_predict_the_future?language=en

[3] W.M. Solomon for the DIII-D Team, "DIII-D Research Advancing the Scientific Basis for Burning Plasmas and Fusion Energy," Nucl. Fusion 57 (2017) 102018.

[4] Zhihong Lin, et al., Turbulent transport reduction by zonal flows: Massively parallel simulations, Science 281 (5384), 1835-1837

[5] Ge Dong, et al., Deep Learning-based Surrogate Model for First-principles Global Simulations of Fusion Plasmas, NUCLEAR FUSION 61 126061 (2021).

[6] David H. Gelernter, Mirror worlds, or, The day software puts the universe in a showbox: how it will happen and what it will mean. Oxford University Press, 1991.

[7] M. Grieves, ''Digital twin: Manufacturing excellence through virtual factory replication,'' NASA, Washington, DC, USA, White Paper 1, 2014.

[8] Aidan Fuller, Zhong Fan, Charles Day, "Digital Twin: Enabling Technologies, Challenges and Open Research", IEEE Access, Vol. 8, 2020.

[9] Jae-Min Kwon, et. al., "Development of a Virtual Tokamak platform", Fusion Engineering and Design, Vol. 184, 2022, 113281.

[10] Summit: www.olcf.ornl.gov (2021).

[11] DOE Incite Program, https://science.osti.gov/ascr/Facilities/Accessing-ASCR-Facilities/INCITE/About-incite

[12] L.L. Lao, et.al. "Reconstruction of current profile parameters and plasma shapes in tokamaks", Nuclear Fusion, Vol. 25, No. 11, 1985.

[13] Julian Kates-Harbeck, Alexey Svyatkosvkiy, William Tang, "Predicting Disruptive Instabilities in Controlled Fusion Plasmas Through Deep Learning", NATURE 568, 526-531 April, 2019.

[14] William Tang, Ge Dong, Jayson Barr, Keith Erickson, Rory Conlin, Dan Boyer, Julian Kates-Harbeck, Kyle Felker, Cristina Rea, N. C. Logan, et al., "Implementation of Ai/Deep Learning Disruption Predictor into a Plasma Control System," arXiv preprint arXiv:2204.01289, 2021; updated version with "Explainable AI/ML Focus" in CONTRIBUTIONS TO PLASMA PHYSICS, Special Issue dedicated to Machine Learning , accepted for publication (April, 2023)

[15] J. L. Barr, et al., "Development and Experimental Qualification of Novel Disruption Prevention Techniques on DIII-D, Nucl. Fusion 61 126019 (2021).

[16] D.M. Jacobsen, R.S. Canon, Contain this, unleashing Docker for HPC, Proceedings of the Cray User Group (2015) https://docs.nersc.gov/development/shifter/files/cug2015udi.pdf

[17] L. Stephey, S. Canon, A. Gaur, D. Fulton, A. J. Younge, "Scaling Podman on Perlmutter: Embracing a community-supported container ecosystem," *2022 IEEE/ACM 4th International Workshop on Containers and New Orchestration Paradigms for Isolated Environments in HPC (CANOPIE-HPC)*, Dallas, TX, USA, 2022, pp. 25-35.

ACKNOWLEDGEMENTS

The authors gratefully acknowledge the collegial contributions of General Atomics in general and Dave Schissel, Sterling Smith, Brian Sammuli, and Raffi Nazikian in particular to the NVIDIA Omniverse visualization of the CAD and EFIT data from DIII-D. Material for the studies presented in this paper is based upon work supported by the U.S. Department of Energy, Office of Science, Office of Fusion Energy Sciences, using the DIII-D National Fusion Facility, a DOE Office of Science user facility, under Awards DE-FC02-04ER54698; DE-AC02-09CH11466, DE-AC52-07NA27344, DE-SC0020337, DE-SC0014264.